\documentclass[prl,aps,showpacs,lengthcheck]{revtex4}
\usepackage{amsmath}
\pdfoutput=1
\usepackage{xcolor}
\usepackage{amsfonts}
\usepackage{graphicx}
\newcommand{\blue}[1]{{\color{blue} #1}}

\newcommand{\green}[1]{{\color{green} #1}}
\begin{document}

\title{Pattern formation in self-propelled particles with density-dependent motility}

\author{F.~D.~C. Farrell$^1$, J. Tailleur$^2$, D. Marenduzzo$^1$,
  M.~C. Marchetti$^3$} \affiliation{$^1$SUPA, School of Physics and
  Astronomy, University of Edinburgh, Mayfield Road, Edinburgh EH9
  3JZ, UK\\ $^2$ Laboratoire Matiere et Systemes complexes, Université
  Paris Diderot, 75205 Paris, France \\ $^3$ Physics Department and Syracuse Biomaterials Institute,
  Syracuse University, Syracuse NY 13244, USA}

\begin{abstract}
We study the behaviour of interacting self-propelled
particles, whose self-propulsion speed decreases with their local
density.  By combining direct simulations of the microscopic model
with an analysis of the hydrodynamic equations obtained by explicitly
coarse graining the model, we show that interactions lead
generically to the formation of a host of patterns, including moving
clumps, active lanes and asters. This general mechanism could explain many of the patterns seen in recent experiments and simulations.
\end{abstract}

\maketitle

Collections of self-propelled (SP) particles
provide the most common realization of active matter, the study of
which constitutes a rapidly growing area of research in
physics~\cite{sriram10}.  Examples of SP particles are bacteria,
cells~\cite{Bray} and actin filaments ``walking'' on a carpet of
immobilized molecular motors~\cite{schaller10}.

The  term ``active'' is used to contrast these systems with their
passive counterparts, such as solutions of diffusing Brownian
particles. Active systems exhibit a much
richer physics than their passive counterparts. 
Most important for us, they have a far larger tendency
to form patterns. For instance, bacterial colonies of e.g. {\it
  E. coli} or {\it S. typhimurium} growing in the lab can
self-organize into crystalline or amorphous arrangements of
high-density bacterial clumps~\cite{murray}, while biofilms form even
more elaborate patterns such as microbial honeycombs, essentially
hexagonal lattices of {\it low-density} spots, or voids~\cite{thar}.
Similarly, actin in high density motility assays~\cite{schaller10} 
organize in moving spots or stripes as well as traveling waves.

What is the mechanism underlying the formation of these ``active
patterns''? One may expect that, as the underlying constituents of
each system are so different, the answer to this question should also
be system-specific. If we are to capture all details of a given active
pattern, this is indeed likely to be the case. Yet, a fascinating
possibility is that there may exist some generic origin of many of
these patterns, stemming from a few universal key features of
activity, linked to its inherent non-equilibrium nature. In some
cases, pursuing such minimal descriptions can be very rewarding. A
well-known example is the hydrodynamic theory of flocking proposed by
Toner and Tu in~\cite{toner98}, which was inspired by the
``agent-based'' model of Vicsek et al.~\cite{vicsek95,sriram10}. The
latter studied the dynamics of an ensemble of SP particles
subjected to aligning interactions, whose ultimate origin may be
hydrodynamic or collision-dominated in the cases of bacteria and actin
filaments, or more complex for bird flocks or fish
schools. Universal features successfully predicted by generic flocking
models are spontaneous motion, giant density fluctuations, and the
emergence of complex spatiotemporal active
patterns~\cite{marchetti09,marchetti10}.

The original Vicsek model considers point particles of fixed speed and includes no interactions between the SP particles other than a rule that aligns their velocities. 
Recently, focus has shifted onto {\it specific} models where additional interactions are included, most commonly steric repulsion~\cite{gregoire2003,peruani06,baskaran08,yang10,marchetti11,peruani11,notestericrepulsion}.  
Our aim here is to develop a more {\it generic} model for interacting SP particles. Interactions are incorporated in our model by assuming that the motility
of the SP particles is a decreasing function of their local density~\cite{note1}.
One may envisage several physical mechanisms responsible for a decay of the propulsion velocity 
with density: here we highlight just two. First, such a slowdown may
arise due to local crowding and steric hindrance, just as
in~\cite{peruani06,baskaran08,peruani11,marchetti11}. An alternative mechanism can be
provided by biochemical signaling such as quorum sensing in bacterial
colonies, as recently explored theoretically~\cite{cates10} and
experimentally~\cite{Hwa11}.  This second mechanism may lead to
slowdown even in dilute suspensions. 
Our work describes the results of simulations of a microscopic  SP
particles model with both interactions and alignment rule, the derivation of the corresponding
hydrodynamic description of the model in terms of a density and a polarization
field, and an analysis of the continuum theory. It therefore provides a direct bridge between microscopic and continuum models, which allows us to identify
universal mechanisms driving pattern formation in interacting SP
particles. As we shall see, interactions lead to an even larger
repertoire of patterns in active particle suspensions than obtained in conventional Vicsek models. These include
moving clumps, lanes and asters, and qualitatively match the patterns
found experimentally, e.g. in~\cite{schaller10}. Our hydrodynamic
equations also provide an understanding of the origin of the various patterns and allow us a
clear comparison with other interacting active particle models.

We consider a modified version of the Vicsek model~\cite{vicsek95},
where the particles interact via a pairwise aligning forcing, which
simplifies the coarse graining of the microscopic model. In 2D the
position $r_i$ and direction, identified by an angle $\theta_i$ (or a vector
$\mathbf{e}_{\theta_i}$), of the
$i$th particle evolve according to
\begin{equation}
{\dot r_i= v\, \mathbf{e}_{\theta_i};\: \dot \theta_i =  \gamma \sum_jF(\theta_j - \theta_i,r_j-r_i) + \sqrt{2\epsilon} \tilde{\eta}_i(t)}
\label{fxy}
\end{equation}
where $\gamma$ and $\epsilon$ are parameters describing the strength
of alignment and fluctuations respectively, and $\tilde{\eta}(t)$ is a
Gaussian white noise with zero mean and unit variance. $F$ controls
the alignment interactions between the spins. For simplicity, we
choose $F(\theta,r)=\sin(\theta)/\pi R^2$ if $|r|<R$ and $0$
otherwise, though its precise shape does not dramatically affect the
physics. \if{As in the Vicsek models, the aligning interactions are
  only computed between particles up to a distance $R$ apart.}\fi In
the $v\to 0$ limit, our model is an off-lattice analogue of the XY
model for a ferromagnet, hence we call it the {\it flying XY
  model}. This differs from other models of SP particles where
alignment is explicitly due to `collisions', and in which the
interaction strength vanishes as $v\to 0$~\cite{bertin09}. Such cases
can however be recovered by taking $\gamma\propto v$.  Last, a
density-dependent velocity is introduced in the model by stipulating
that $v$ depends on the number of particles $n$ within a given radius
$R_n$, as $v(n) = v_0 e^{-\lambda n} + v_1$, where $v_0\gg v_1>0$ are
the dilute and crowded limiting velocities respectively, and
$\lambda>0$ controls the decay of the motility decreases with increasing
density. Hereupon we restrict to $R_n=R$.

\begin{figure}
\includegraphics[width=0.49\columnwidth]{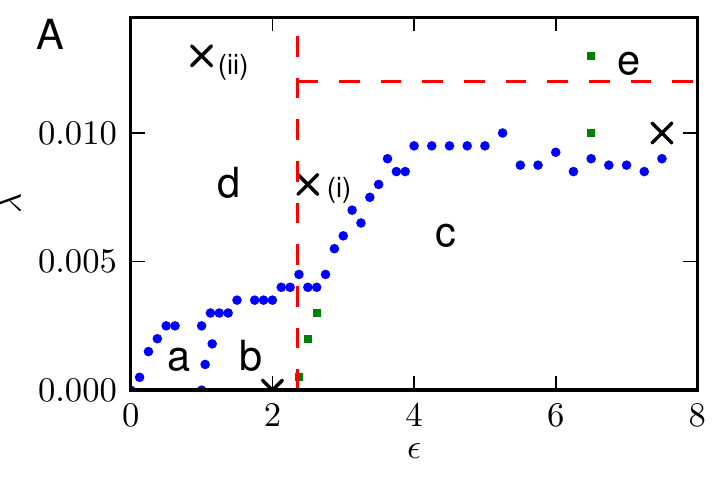}
\includegraphics[width=0.49\columnwidth]{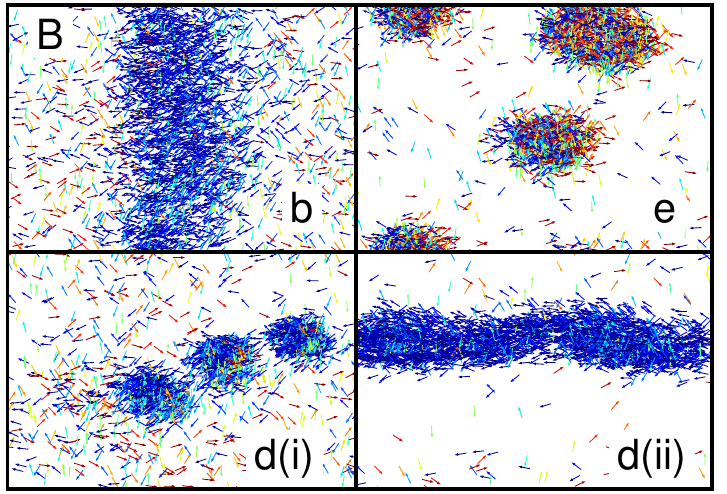}
\caption{(color online) (A) Phase diagram in the $(\epsilon,\lambda)$ plane, for
$N=3000$, $L=10$, $\gamma=0.16$, $v_0=2$ and $v_1=0.1$. 
Blue filled circles on the phase boundary correspond to 
peaks in the variance of the particle density, 
while green squares separate states with zero mean orientation from
states with nonzero mean orientation. 
Phases are labelled as per discussion in the text. 
Horizontal and vertical red lines indicate linear instabilities towards 
clustering and ordering, respectively.  (B) Snapshots of the stripy (b), aster (e), moving clumps (d(i)), lane (d(ii)) patterns. The crosses in A correspond to
the snapshots in B. Particles are color coded by direction, with
blue horizontal and red vertical.
}
\label{phasediagram}
\end{figure}

Fig.~\ref{phasediagram}A shows a representative phase diagram of the
flying XY model in the $(\epsilon,\lambda)$ plane.  For small
$\lambda$, when $v$ is practically constant, the phases observed are
the same as those in the literature on flocking
models~\cite{vicsek95,chate04}. Namely, at high $\epsilon$ we find a
disordered, homogeneous state (region c in Fig. 1A), followed by a
polarly ordered phase with high density stripes (named stripy phase, b
in Fig. 1A) below a critical noise value. For even lower $\epsilon$,
we observe a `fluctuating flocking state' (region a) with polar
order and large density fluctuations -- this state is very close to
the one described in Ref.~\cite{marchetti10} and we do not discuss it
further here.  These phases are expected -- as we shall see below the
hydrodynamic equations for our model map to those for the Vicsek
model~\cite{toner98,marchetti10} when $v$ is constant.

 Above a critical value $\lambda_c(\epsilon)$, new patterns
 appear. Due to the density-dependent motility, the SP particles
 cluster via a self-trapping mechanism through which they assemble and
 slow-down, creating a positive feedback loop akin to the one
 discussed for models of bacteria~\cite{cates10}. This process leads
 to the formation of high density clumps which slowly coarsen towards
 a fully phase separated steady state. The Vicsek-like alignment tendency 
 greatly affects this instability. On one hand, the critical value
 $\lambda_c(\epsilon)$ decreases almost to zero with 
decreasing $\epsilon$. Furthermore,
 the presence of polar order promoted by the alignment changes the
 nature of the clusters. In Fig. 1A we identify at least three
 distinct patterns, of which snapshots are shown in Fig. 1B. When
 $\epsilon$ is small, rather than structureless dots, the clusters
 show an orientational order and move coherently: they form ``moving
 clumps'' (pattern d(i) in Fig. 1). For low $\epsilon$ and large
 $\lambda$ the moving clumps merge into bands, or lanes (labelled as
 d(ii)) -- within these, however, particles move {\it parallel} rather
 than perpendicular to the band, in contrast with the $\lambda\to 0$
 stripy phase.  
This is reminiscent of the `streaks' of actin
 filaments observed in~\cite{schaller10}. In the disordered, high
 $\epsilon$ phase, the clusters instead diffuse randomly, and are on
 average stationary. 

Here, a temporal average of the particle
 orientation patterns shows that the clusters are asters (the aster
 phase is labelled as e in Fig. 1). However, as discussed in greater
 detail below, the orientation in the aster is non-standard:
 particles point towards its center at the core, but they
 coherently point outwards in its periphery.  We stress that
 moving clusters, lanes and asters are not observed either in the
 standard Vicsek model, or in its standard mean field continuum
 description used in Ref.~\cite{marchetti10}. They are a consequence
 of the interplay between self-propulsion, alignment and
 density-dependent motility: switching off any of these ingredients
 would thus greatly reduce the pattern forming potential of the
 model.

To get a better understanding of the pattern formation process, we now
discuss how to coarse grain the microscopic dynamics~\eqref{fxy} to
obtain a macroscopic description of the model. They are two candidates for
the hydrodynamic fields: the particle density $\rho$, which is
conserved, and the local alignment, or polarization, vector ${\mathbf
  P}$, on symmetry ground. Note that ``hydrodynamic'' here means
slowly varying in space and time -- the dynamics of the underlying
fluid is not included in our modeling.  Following Ref.~\cite{dean96}, 
we start with the microscopic
Eq.~\eqref{fxy} and use It\=o calculus~\cite{cates10} 
to write down a stochastic dynamical
equation for the evolution of $f(\mathbf{r},\theta,t)$, the density of
particles at position $\mathbf{r}$ with angle $\theta$, as follows
\begin{equation}
\label{coarsegrained}
\begin{aligned}
  &\partial_t f(\mathbf{r},\theta) + \mathbf{e}_{\theta} \cdot \nabla[v f]   = \epsilon \frac{\partial^2 f}{\partial \theta^2} - \frac{\partial}{\partial \theta} \sqrt{2\epsilon f}\eta \\ 
& -\gamma \frac{\partial}{\partial \theta} \int  d\theta' d {\mathbf r'} f(\mathbf{r'},\theta')f(\mathbf{r},\theta)F(\theta'-\theta,\mathbf{r}-\mathbf{r'}).
\end{aligned}
\end{equation}
The second term on the left hand side describes familiar
advection, but with one important difference: the velocity $v$ appears {\it inside} the gradient.
This is precisely what leads to the instabilities
responsible for the new patterns in the simulations. By
dropping the conserved noise $\sqrt{2\epsilon f}\eta$, we obtain a
non-fluctuating kinetic equations for the flying XY model. Following
Bertin et al.~\cite{bertin09}, we Fourier transforms
Eq.~\ref{coarsegrained} to get equations of motion for $f_k\equiv\int
f e^{i k \theta} d\theta$. Using $2\pi f(\theta)=\sum_k f_k e^{-ik\theta}$
and $2\pi F(\theta)=\sum_k F_k e^{-ik\theta}$, we obtain a
hierarchy of equations:
\begin{equation}
  \label{eqn:FT}
  \begin{aligned}
    &\partial_t f_k +  \frac{\partial}{\partial x} \frac{v f_{k+1} + v f_{k-1}}2  + \frac{\partial}{\partial y} \frac{v f_{k+1} - v f_{k-1}}{2 i} \\  
& = -k^2 \epsilon f_k + i \frac{\gamma k}{2\pi} \sum_m f_m F_{-m} f_{k-m},
  \end{aligned}
\end{equation}
where all sums run from $-\infty$ to $+\infty$. In principle, $F$ is
slightly non-local in space so that the second term of the r.h.s. of
Eq.~\eqref{eqn:FT} should retain a spatial integral. We are
however interested in the hydrodynamic, large-scale, description of
the system, a limit in which $R$ is very small and we assume $F$ to be
perfectly local~\cite{note2}.  To obtain mean
field equations for the hydrodynamic variables, we note that $f_k$ for
$k=0$ is simply the density, $\rho$, whereas the real and imaginary
part of $f_1$ are the $x$ and $y$ component of $\rho \mathbf{P}$
respectively (as we work in 2D we can identify vectors with complex
numbers). By writing out in full the $k=0$ case of
Eq.~\eqref{coarsegrained}, we then find that the density field obeys the
continuity equation
\begin{equation}
\partial_t \rho = - \nabla \cdot (v  \mathbf{W}) \label{density},
\end{equation} 
where ${\mathbf W}\equiv \rho{\mathbf P}$.
To make further progress, we now assume that we are not too deeply in
the ordered phase, so that $f(\theta)$ is to first order approximation
homogeneous, hence higher Fourier components ($f_k$ for $k \ge 3$) may
be neglected. Following~\cite{bertin09}, we further assume that $f_2$
is a fast variable, so that $\dot{f_2} \simeq 0$. After lengthy but
straightforward algebra, we obtain the following
equation for ${\mathbf W}$,
\begin{equation}
   \begin{aligned}
    & \partial_t \mathbf{W} +\frac{\gamma}{16 \epsilon}(\mathbf{W} \cdot \nabla)(v \mathbf{W}) =   (\frac{1}{2} \gamma \rho - \epsilon)\mathbf{W} \\
    & - \frac{\gamma^2}{8 \epsilon}W^2 \mathbf{W}  - \frac{1}{2}\nabla(v\rho) + \frac{3\gamma}{16 \epsilon}\nabla(v W^2) \\
    & - \frac{\gamma}{32 \epsilon}v\nabla W^2 - \frac{3\gamma}{16 \epsilon}\mathbf{W} \nabla \cdot (v \mathbf{W}) \\
 &-\frac{\gamma}{8 \epsilon}v\mathbf{W} (\nabla \cdot \mathbf{W}) 
-\frac{\gamma}{8 \epsilon}v(\mathbf{W} \cdot \nabla) \mathbf{W}+{\cal O}(\nabla^2) 
  \label{hydro}  \end{aligned}
\end{equation}

The second term on the l.h.s. of Eq.~\eqref{hydro} describes
self-advection of particles and breaks Galilean
invariance~\cite{toner98}.  The first two terms on the right-hand side
describe the standard spontaneous symmetry breaking leading to polar
order and flocking for sufficiently small $\epsilon$ in the Vicsek
model at $\lambda=0$. The third, pressure-like term,
$-\frac{1}{2}\nabla(v\rho)$, is the most relevant one in our work, as
it is responsible for the clustering instability observed in Fig. 1
when $\lambda\ne 0$. Higher order terms in $\nabla$ and $\mathbf W$
have minor effects on patterns and will be discussed
elsewhere.

Having written down the mean field equations of motion,
Eqs.~\eqref{density} and \eqref{hydro}, we can now assess how their
predictions compare with the full simulations of the microscopic
model. 
The continuum theory predicts an order-disorder transition at 
$\epsilon_c=\frac{1}{2} \gamma \rho_0$. For $\epsilon>\epsilon_c$ there is a stable homogeneous disordered state, with $\rho=\rho_0$ and 
$\mathbf{W}=0$.   For $\epsilon<\epsilon_c$ the equations yield a homogeneous ordered or flocking state with
$\rho=\rho_0$ and 
$\mathbf{W}=W_0{\bf \hat{x}}$, where we have chosen the $x$ axis along the direction of broken symmetry and 
$W_0=\sqrt{8\epsilon(\epsilon_c-\epsilon)/\gamma^2}$. The mean-field transition at $\epsilon_c$ does not depend on $\lambda$ and coincides with that of the {\it equilibrium} XY model. The order-disorder phase boundary predicted by the theory is compared to its numerical counterpart in Fig. 2A.  
We then study the linear stability of the homogeneous disordered state at $\epsilon>\epsilon_c$ against spatially inhomogeneous fluctuations. It is straightforward to show that when $\lambda\ne 0$  the homogeneous disordered phase
becomes unstable for all wavenumbers when $v(\rho_0) + \rho_0
v'(\rho_0) < 0$. This instability, referred to as a clustering instability, arises due to the term -
$\frac{1}{2}\nabla(v\rho)$ in the equation for $\mathbf{W}$. The
threshold between homogeneous and clustered phases found numerically
at large $\epsilon$ is close to but below the prediction
(Fig. 2B). This is reasonable, as the linear stability can only access
the spinodal line: fluctuations may trigger phase separation for lower
$\lambda$.  
\if{
\blue{Similarly we study the  stability of the homogeneous ordered state for $\epsilon<\epsilon_c$ by considering the linear dynamics of fluctuations $\delta\rho({\bf r},t)=\rho-\rho_0$ and $\delta\mathbf{W}({\bf r},t)=\mathbf{W}-\mathbf{W}_0$. The analysis of the mode structure will be given elsewhere. We find ... \green{I think we should do this calculation.I expect two instabilities here: the one found by Bertin et al (and Mishra et al), although probably modified by $lambda$, and then another one associated with the negative derivative of $v(\rho)$. The latter I would call clustering instability of the ordered state and it may occur at a lower value of $\lambda$ than the clustering instability of the disordered state.} Finally, in Fig. 1A the vertical dashed line indicates the mean-field transition between homogeneous disordered and ordered states, while the horizontal dashed line denotes the clustering instability of the homogeneous disordered state.}
}
\fi
\begin{figure}
\centering
\includegraphics[width=0.49\columnwidth]{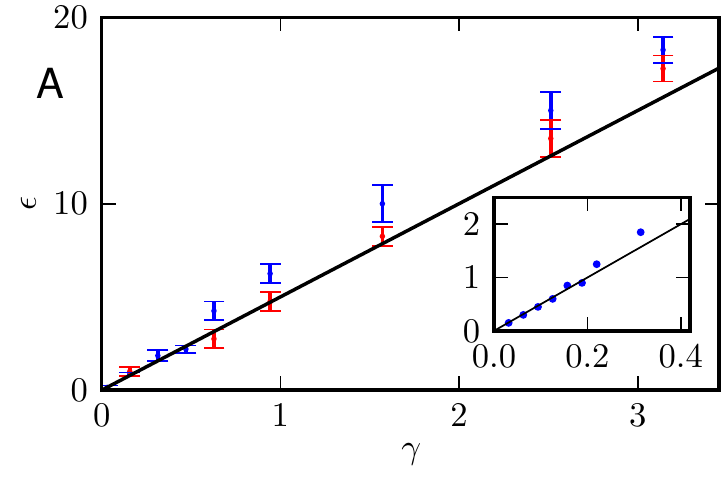}
\includegraphics[width=0.49\columnwidth]{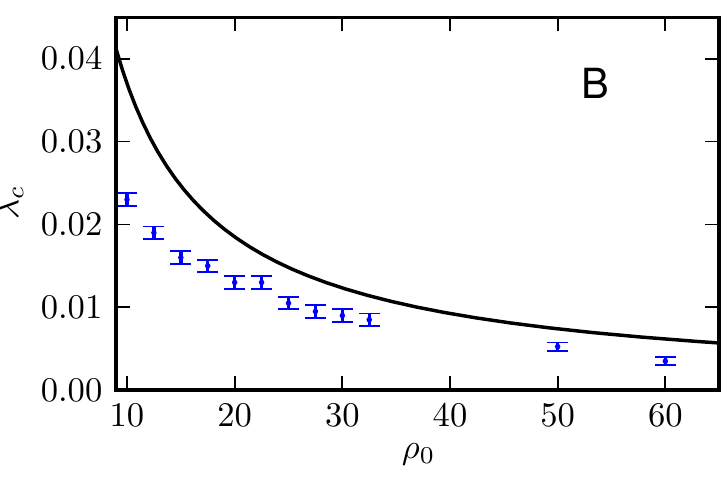}
\vspace{-20pt}
\caption{(color online) (A) Phase boundary for the flying XY model when $\lambda=0$,
  showing the critical value of $\epsilon$ as a function of
  $\gamma$. Blue points for $v=2.0$, red for $v=0.5$. Inset: data for
  $v=2.0$ for smaller values of $\gamma$, showing good agreement. (B)
  Phase boundary for $\epsilon=5$, $\gamma=0.16$. In all cases $L=10$
  and $N=1000$.}
\label{phaseboundary}
\end{figure}


To go beyond the simple linear stability analysis of the homogeneous disordered state, account for the effect of the non-linear terms, and hence explore
the range of patterns compatible with our hydrodynamics equations, we
solved Eqs.~\eqref{density} and \eqref{hydro} numerically, by means of
a standard finite difference scheme. In order to enhance the stability
of our algorithm, we included a diffusive term $D \nabla^2 \rho $ on
the right hand side of Eq.~\eqref{density}. 
Our numerical results show that all the
five patterns, or phases, observed in the microscopic simulations
(fluctuating flocking state, moving stripes and lanes, static asters
and moving clumps) can be found within Eqs.~\eqref{density} and
\eqref{hydro} -- Fig. 3 portrays a comparison of the $\lambda\ne 0$
patterns. Interestingly the origin of the atypical asters can be
directly read in Eq.~\eqref{hydro}. In the steady-state, low gradient,
small ${\mathbf{W}}$ approximation, Eq.~\eqref{hydro} reduces to
$(\gamma\rho/2-\epsilon)\mathbf W=\nabla(\rho v/2)$ and
$\nabla(v\rho)$ thus acts as an ordering field for ${\mathbf
  W}$. Along the radius of an aster, the density increases towards the
center whereas the velocity decreases. Their product can thus be
non-monotonous, which makes $\mathbf W$ change direction, whence the
atypical asters seen in the microscopic simulations. In the continuum
simulations, even though $\nabla \rho v$ can change sign, the presence
of the diffusive terms disallows sharp gradients and we did
not find parameters for which $\nabla \rho v$ was dominating. We
could, however, end up with both inward pointing or outward pointing
asters, corresponding to phases with high-density clumps (at small
$\lambda$, shown in Fig.~3C) or low-density voids (at larger
$\lambda$, similar to those discussed in~\cite{thar}, not shown).

\begin{figure}
  \label{patterns}
  \begin{minipage}[l]{.7\columnwidth}
    \hspace{-40pt}
    \includegraphics[width=\columnwidth]{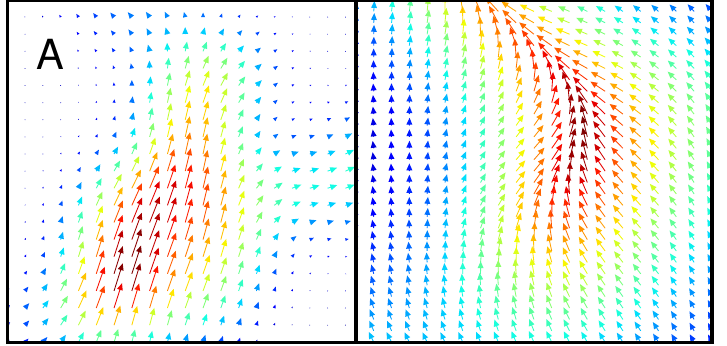}
  \end{minipage}
  \hspace{-25pt}
  \begin{minipage}[l]{0.25\columnwidth}
    \resizebox{2cm}{!}{
      \begin{tabular}{l | c | r}
        $\tilde{\lambda}$ & 0.75 & 1.5 \\
        $\tilde{v_1}$ & 0.05 & 0.09 \\
        $\tilde{\gamma}$ & 1.6 & 3.6 \\
        $\tilde{D}$ & - & 0.02 \\ 
      \end{tabular}
    }
  \end{minipage}
  
  \begin{minipage}[l]{0.7\columnwidth}
    \hspace{-40pt}
       \includegraphics[width=\columnwidth]{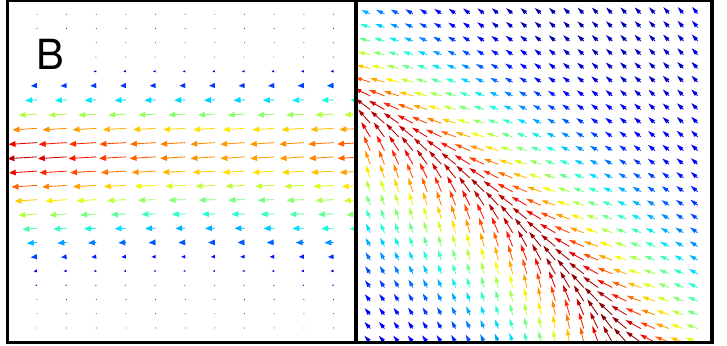}
  \end{minipage}
      \hspace{-25pt}
      \begin{minipage}[r]{0.25\columnwidth}
      \resizebox{2cm}{!}{
        \begin{tabular}{l | c | r}
        $\tilde{\lambda}$ & 1.2 & 1.5 \\
        $\tilde{v_1}$ & 0.05 & 0.12 \\
        $\tilde{\gamma}$ & 4.8 & 4.5 \\
        $\tilde{D}$ & - & 0.02 \\ 
        \end{tabular}
      }
  \end{minipage}
  
  \begin{minipage}[l]{0.7\columnwidth}
    \hspace{-40pt}
    \includegraphics[width=1\columnwidth]{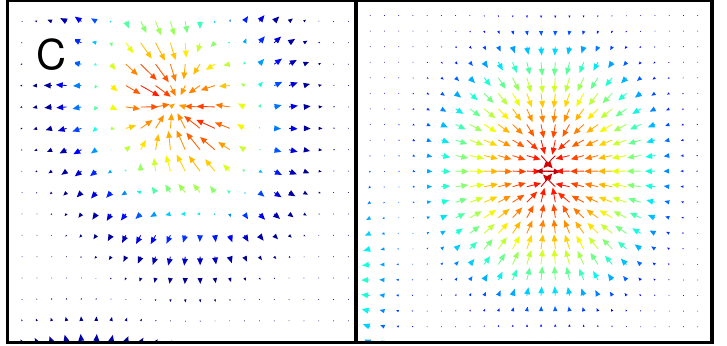}
  \end{minipage}
      \hspace{-25pt}
      \begin{minipage}[r]{0.25\columnwidth}

      \resizebox{1.9cm}{!}{
        \begin{tabular}{l | c | r}
        $\tilde{\lambda}$ & 0.94 & 1.5 \\
        $\tilde{v_1}$ & 0 & 0 \\
        $\tilde{\gamma}$ & 0.64 & 2.3 \\
        $\tilde{D}$ & - & 0.01 \\ 
      \end{tabular}}
  \end{minipage}
  
  \caption{(color online) Patterns found for $\lambda\ne 0$ in the microscopic
    simulations (left column) and in the numerical solution of the
    hydrodynamic equations (right). Tables show dimensionless parameter values: $\tilde{\lambda}=\lambda\rho_0$, $\tilde{v_1}=v_1/v_0$, $\tilde{\gamma}=\gamma\rho_0/\epsilon$, $\tilde{D}=D\epsilon/v_0^2$. Arrows show the ${\mathbf W}$ field, colors the density (red:
    high; blue: low).}
\end{figure}

We have shown that  a density-dependent motility
in our flying XY model, which is a close relative of the Vicsek model, yields
new patterns in suspensions of SP particles. Such patterns
include moving clumps, lanes, and asters, associated with high-density
clumps, or voids. All these patterns have experimental
counterparts~\cite{schaller10,murray,thar}. By explicitly linking the
microscopic and coarse grained mean field dynamics, we were able to identify 
the key ingredients which trigger the appearance of the new patterns
in the
``pressure term'' $-\frac{1}{2}\nabla(v\rho)$: when this turns
negative, new patterns may form. Importantly, the patterns we see are
not very sensitive to the precise form of $v(\rho)$. For instance,
steric hindrance results in velocities that typically decrease
linearly with density~\cite{thompson2011} and would give  similar
instabilities.

We close with a comparison with other models featuring patterns
similar to ours. Continuum models for microtubule-kinesin solutions
leading to aster formation have been proposed in~\cite{leekardar}.
The resulting equations of motion for the microtubule polarisation
included a phenomenological term of the form $S\nabla(\rho)$ with
$S>0$, and $\rho$ the density of motors bound to microtubules. This is
qualitatively similar to our equations with negative
$-\frac{1}{2}\nabla(v\rho)$. In the $\lambda=0$ limit,
Refs.~\cite{bertin09,marchetti10} show that asters are absent if the
prefactors in the non-linear terms in the continuum equations are obtained via a systematic coarse-graining of a
system of Vicsek SP particles or SP hard rods. {Ref.~\cite{gopinath} shows, however, that asters do appear if these prefactors are tuned as independent parameters, although in this case the asters have fixed polarity.} 
\if{This work also identifies
 the change in sign of the effective compressibility of the system as the key mechanism controlling pattern formation.}\fi

Finally, Peruani {\it et al.}~\cite{peruani11} studied a microscopic 
lattice variant of the
Vicsek model, and also found asters and moving clumps, dubbed traffic
jams and gliders respectively. The underlying passive model in
Ref.~\cite{peruani11} is the 4-state Potts model, which is in a
different universality class than the XY model, to which our dynamics
reduces in the passive limit. However, the {\it active} patterns are
similar in the two models. This is again naturally explained by our
theory, as their origin in~\cite{peruani11} lies in the slowdown of
particles due to crowding jamming, which brings up an effective
``pressure term'' analogous to the one in Eq.~\eqref{hydro}. A
density-dependent motility, induced either by steric hindrance or by
crosslinkers between actin filaments, may also at the basis of the
formation of similar patterns in the actin-walker experiments
in~\cite{schaller10}. 

We thank M.~R. Evans for useful discussions.
MCM was supported by the National Science Foundation through awards
DMR-0806511 and DMR-1004789.


\begin{thebibliography}{99}
\bibitem{notestericrepulsion} One should further distinguish between collection of sterically interacting SP particles where alignment is imposed a-la-Vicsek, and others were it comes solely from steric repulsion of rod-like particles~\cite{peruani06,baskaran08,yang10,peruani11}.  In the latter case steric repulsion aligns SP particles regardless of their polarity, yielding nematic rather than polar order.
\bibitem{note1} Other
  functional dependencies lead to a less rich physics -- details will
  be given elsewhere.
\bibitem{note2} A non-local $F$ can be dealt e.g. as in~\cite{cates10}.
\end{thebibliography}
\end{document}